\documentclass[12pt,epsf]{article}
\usepackage{graphicx,amsmath,amssymb}
\begin{document}
%%%%%%%%%%%%%%%%%%%%%%%%%%%%%%%%%%%%%%%%%%%

\def\a{\alpha}
\def\b{\beta}
\def\c{\varepsilon}
\def\d{\delta}
\def\e{\epsilon}
\def\f{\phi}
\def\g{\gamma}
\def\h{\theta}
\def\k{\kappa}
\def\l{\lambda}
\def\m{\mu}
\def\n{\nu}
\def\p{\psi}
\def\q{\partial}
\def\r{\rho}
\def\s{\sigma}
\def\t{\tau}
\def\u{\upsilon}
\def\v{\varphi}
\def\w{\omega}
\def\x{\xi}
\def\y{\eta}
\def\z{\zeta}
\def\D{\Delta}
\def\G{\Gamma}
\def\H{\Theta}
\def\L{\Lambda}
\def\F{\Phi}
\def\P{\Psi}
\def\S{\Sigma}

\def\o{\over}
\def\beq{\begin{eqnarray}}
\def\eeq{\end{eqnarray}}
\newcommand{\gsim}{ \mathop{}_{\textstyle \sim}^{\textstyle >} }
\newcommand{\lsim}{ \mathop{}_{\textstyle \sim}^{\textstyle <} }
\newcommand{\vev}[1]{ \left\langle {#1} \right\rangle }
\newcommand{\bra}[1]{ \langle {#1} | }
\newcommand{\ket}[1]{ | {#1} \rangle }
\newcommand{\EV}{ {\rm eV} }
\newcommand{\KEV}{ {\rm keV} }
\newcommand{\MEV}{ {\rm MeV} }
\newcommand{\GEV}{ {\rm GeV} }
\newcommand{\TEV}{ {\rm TeV} }
\def\diag{\mathop{\rm diag}\nolimits}
\def\Spin{\mathop{\rm Spin}}
\def\SO{\mathop{\rm SO}}
\def\O{\mathop{\rm O}}
\def\SU{\mathop{\rm SU}}
\def\U{\mathop{\rm U}}
\def\Sp{\mathop{\rm Sp}}
\def\SL{\mathop{\rm SL}}
\def\tr{\mathop{\rm tr}}

\def\IJMP{Int.~J.~Mod.~Phys. }
\def\MPL{Mod.~Phys.~Lett. }
\def\NP{Nucl.~Phys. }
\def\PL{Phys.~Lett. }
\def\PR{Phys.~Rev. }
\def\PRL{Phys.~Rev.~Lett. }
\def\PTP{Prog.~Theor.~Phys. }
\def\ZP{Z.~Phys. }

%%%%%%% added by Fumi %%%%%%%%%%
% FROM HERE
%\newcommand{\beq}{\begin{equation}}   
%\newcommand{\eeq}{\end{equation}}
\newcommand{\bea}{\begin{eqnarray}}   
\newcommand{\eea}{\end{eqnarray}}
\newcommand{\bear}{\begin{array}}  
\newcommand {\eear}{\end{array}}
\newcommand{\bef}{\begin{figure}}  
\newcommand {\eef}{\end{figure}}
\newcommand{\bec}{\begin{center}}  
\newcommand {\eec}{\end{center}}
\newcommand{\non}{\nonumber}  
\newcommand {\eqn}[1]{\beq {#1}\eeq}
\newcommand{\la}{\left\langle}  
\newcommand{\ra}{\right\rangle}
\newcommand{\ds}{\displaystyle}
\def\SEC#1{Sec.~\ref{#1}}
\def\FIG#1{Fig.~\ref{#1}}
\def\EQ#1{Eq.~(\ref{#1})}
\def\EQS#1{Eqs.~(\ref{#1})}
\def\TEV#1{10^{#1}{\rm\,TeV}}
\def\GEV#1{10^{#1}{\rm\,GeV}}
\def\MEV#1{10^{#1}{\rm\,MeV}}
\def\KEV#1{10^{#1}{\rm\,keV}}
\def\lrf#1#2{ \left(\frac{#1}{#2}\right)}
\def\lrfp#1#2#3{ \left(\frac{#1}{#2} \right)^{#3}}
\def\REF#1{Ref.~\cite{#1}}
\newcommand{\osc}{{\rm osc}}
\newcommand{\ed}{{\rm end}}
\def\dda#1{\frac{\partial}{\partial a_{#1}}}
\def\ddat#1{\frac{\partial^2}{\partial a_{#1}^2}}
\def\dd#1#2{\frac{\partial #1}{\partial #2}}
\def\ddt#1#2{\frac{\partial^2 #1}{\partial #2^2}}
\def\lrp#1#2{\left( #1 \right)^{#2}}
% UNTIL HERE

%%%%%%%%%%%%%%%%%%%%%%%%%%%%%%%%%%%%%%%%%%%%%%%%%%%%%%%%%%%%%%%%%%%

\baselineskip 0.7cm

\begin{titlepage}

\begin{flushright}
UT-14-25\\
TU-971\\
IPMU14-0122\\
\end{flushright}

\vskip 1.35cm
\begin{center}
{\large \bf 
Extra light fermions in  $E_6$-inspired models and the 3.5 keV X-ray line signal
}
\vskip 1.2cm
Kazunori Nakayama$^{a,c}$,
Fuminobu Takahashi$^{b,c}$
and 
Tsutomu T. Yanagida$^{c}$

\vskip 0.4cm

{\it $^a$Department of Physics, University of Tokyo, Tokyo 113-0033, Japan}\\
{\it $^b$Department of Physics, Tohoku University, Sendai 980-8578, Japan}\\
{\it $^c$Kavli Institute for the Physics and Mathematics of the Universe (WPI), TODIAS, University of Tokyo, Kashiwa 277-8583, Japan}

\vskip 1.5cm

\abstract{
We propose a scenario in which extra light fermions in  an $E_6$-inspired U(1) extension of the standard model
constitute the dark matter, as a simple variation of our model for dark radiation presented in 2010.
Interestingly, for the light fermions of mass  about $7$\,keV, they radiatively decay into active neutrinos and photons 
with a lifetime in the range of $10^{27}-10^{28}$ seconds, which naturally explains
the recently discovered $3.5$\,keV X-ray line signal. 
}
\end{center}
\end{titlepage}

\setcounter{page}{2}

%%%%%%%%%%%%%%%%%%%%%%%%%%%
%\section{Introduction}
%%%%%%%%%%%%%%%%%%%%%%%%%%%

Recently two groups independently reported detection of an unidentified X-ray line at about $3.5$~keV in the XMM-Newton 
X-ray observatory data of various galaxy clusters and the Andromeda galaxy~\cite{Bulbul:2014sua,Boyarsky:2014jta}.
The observed X-ray line could be due to decaying dark matter (DM), although there are a variety of systematic uncertainties 
that might affect the observed line energy and flux. 
In this Letter we pursue a possibility that decaying dark matter of mass about $7$\,keV explains the excess of the X-ray line
at $3.5$\,keV.

One of the plausible candidates for light decaying dark matter is sterile neutrinos~\cite{Dodelson:1993je,Dolgov:2000ew,Abazajian:2001vt}. See 
Refs.~\cite{Boyarsky:2009ix,Kusenko:2009up,Abazajian:2012ys} for a review on sterile neutrino dark matter and 
its various implications. Sterile neutrinos radiatively decay into neutrinos and photons via their mixings with the active neutrinos.
The mass and the lifetime of the sterile neutrino dark matter preferred by the observed X-ray line excess 
are~\cite{Bulbul:2014sua,Boyarsky:2014jta}.
%These excess events are suitably explained by decaying dark matter (DM) models with mass and lifetime
%%
\begin{gather}
	\label{mass}
	m_{\rm DM} \simeq 7\,{\rm keV}, \\
	\tau_{\rm DM} \simeq 2\times 10^{27} - 2\times 10^{28}\,{\rm sec},
	\label{tau}
\end{gather}
where the total dark matter is assumed to be made of the sterile neutrinos.
Such extremely light sterile neutrino together with other two heavy right-handed neutrinos 
can be realized in an extra dimensional set-up~\cite{Kusenko:2010ik}, while retaining 
the seesaw formula~\cite{seesaw}. It has been known, however, that such light sterile neutrino should not contribute to
the light neutrino mass through the seesaw mechanism in order to satisfy the existing X-ray bounds~\cite{Asaka:2005an,Boyarsky:2006jm}.
Indeed,   the lifetime would be several orders of magnitude  shorter than the required value (\ref{tau}) for the sterile 
neutrino mass $7$\,keV~\cite{Ishida:2014dlp}, which signals the necessity for some extension such as extra flavor structure.\footnote{
Another possibility is to introduce a set of chiral fermions charged under U(1)$_{\rm B-L}$, where the number and the charge assignment 
are tightly constrained by the anomaly cancellation, and at least five additional chiral fermions are necessary~\cite{Nakayama:2011dj}.
For one of the solutions, the lightest chiral fermion becomes naturally as light as ${\cal O}({\rm keV})$ and may decay into neutrinos and 
photons if $Z_{2}({\rm B-L})$ is approximate. 
} 
% Thus, it is of crucial importance to understand the required longevity of the sterile neutrino dark matter. 
 In particular, the fact that the sterile neutrino dark matter should not contribute to the light neutrino mass
 through the seesaw formula suggests that it might have a different origin from that of the heavy right-handed neutrinos. 
 
In this short Letter we propose an alternative to the right-handed sterile neutrino dark matter,
 as the origin of the $3.5$\,keV X-ray line. 
We explore the possibility that extra light fermions  in an extension of the standard model (SM) 
gauge group become dark matter which decays into photons explaining the observed X-ray line excess. 
Intriguingly, we shall see that the lifetime falls in the right range of (\ref{tau}) for the given dark matter mass of  $7$\,keV.

Let us consider an $E_6$-inspired extension of the SM, based on the model for dark radiation proposed by
the present authors in 2010~\cite{Nakayama:2010vs}.
Here we consider a gauge group of two additional anomaly-free U(1)s, ${\rm SU(5)} \times {\rm U(1)}_\psi \times
{\rm U(1)_{B-L}}$,  inspired by the $E_6$ model of the GUT~\cite{Langacker:2008yv}.\footnote{
Here the charge assignment of the U(1)$_{\rm B-L}$ is given by a certain combination of the U(1) hypercharge and
 the ${\rm B-L}$. } Here
SU(5) includes the SM gauge groups.  The $E_6$ group has a ${\bf 27}$ representation, which can be decomposed as
${\bf 27 = (10+{\bar 5}+1)_1 + (5+{\bar 5})_{-2} +1_{4}}$ in terms of the SU(5) representation,
where the subscript denotes the U(1)$_\psi$ charge. See Table~\ref{table} for notation and charge assignments.
 All the SM fermions plus a right-handed neutrino in one generation are contained in the first multiplet.
On the other hand, there are extra fermions, $\psi_{\bf 5}$, $\psi_{\bf \bar 5}$ and $\psi_{\bf 1}$,  in the ${\bf 27}$ representation,
where the subscript indicates the representation under SU(5). The SM gauge singlet fermion, $\psi_{\bf 1}$, is 
a candidate for dark radiation or light dark matter~\cite{Nakayama:2010vs}.

%%%%%%%%%%%%%%%% table %%%%%%%%%%%%%%%%%%%%%%
\begin{table}[t]
  \begin{center}
    \begin{tabular}{| c | c | c | c |}
      \hline 
         SU(5) & U(1)$_\psi$  & U(1)$_{\rm B-L}$ & $Z_2$ \\
       \hline 
       $\psi_{\bf 10}^{\rm (SM)}$         & 1 &    1     & even\\
       $\psi_{\bf \bar 5}^{\rm (SM)}$   & 1 &  $-3$ & even \\
       $\psi_{\bf 1}^{\rm (SM)}\equiv\nu_{\rm R}$  & 1 & 5 & even \\
      \hline
       $\psi_{\bf 5}$                              &$-2$ & $-2$ & odd \\
       $\psi_{\bf \bar 5}$                      &$-2$ & $2$  & odd \\
      \hline 
       $\psi_{\bf 1}\equiv \psi$         &4       &0       & even \\
       \hline\hline
      $\phi_{\bf 1}\equiv \phi$           & 1       & 5      & even\\
      \hline
       $\phi_{\bf 5}\supset H$            &$-2$  &$-2$ & even\\
       $\phi_{\bf \bar 5}\supset \bar H$    &$-2$  &$2$ & even\\
      \hline 
       $X_{\bf 1}\equiv X$              & $n$       & 0      & even \\
        \hline
    \end{tabular}
    \caption{ 
    		Notation and charge assignments on the fields in the $E_6$-inspired model.
           }
    \label{table}
  \end{center}
\end{table}
%%%%%%%%%%%%%%%%%%%%%%%%%%%%%%%%%%%%%%%%%%%%%% 

We introduce two SM-singlet Higgs fields: $\phi$ and $X$, whose vacuum expectations values (VEVs)
induce the spontaneous breaking of the two U(1)s.  See Table~\ref{table} for their charge assignments.
First, the VEV of $\phi$ breaks a combination of two U(1)s, giving rise to the right-handed neutrino mass
as
\begin{equation}
	\mathcal L = \frac{1}{2M}\la \phi^{*} \ra^2\nu_{\rm R} \nu_{\rm R} + {\rm h.c.},
\end{equation}
where $M$ denotes the cutoff scale. The right-handed neutrino mass will be
about $\GEV{15}$ for $\langle \phi\rangle \sim 5\times10^{16}\,$GeV and  $M =  M_P \simeq 2.4\times 10^{18}\,$GeV, and
the light (active) neutrino masses can be explained by the seesaw mechanism~\cite{seesaw}.
Note that the other combination, $U(1)_X \equiv 5U(1)_\psi - U(1)_{\rm B-L}$, remains unbroken, and
fermions except for $\nu_{\rm R}$ do not get masses by the VEV of $\phi$. We also impose a
$Z_2$ parity (see Table~\ref{table}) to forbid
a mixing between extra colored fermion and SM fermions, $\mathcal L  = \phi \psi_{\bf \bar 5}^{\rm (SM)} \psi_{\bf 5}+{\rm h.c.}$.
Also the interactions with the Higgs fields $\phi_{\bf 5}$ and $\phi_{\bar {\bf 5}}$ and the extra fermions,
$\phi_{\bf 5} \psi_{\bar {\bf 5}} \psi_{\bf 1}$ and $\phi_{\bf {\bar 5}} \psi_{{\bf 5}} \psi_{\bf 1}$, are forbidden.

Secondly, the remaining U(1)$_X$ is broken by the VEV of the other SM-singlet Higgs, $X$,
whose U(1)$_\psi$ charge is represented by $n$. If $n=4$, both $\phi$ and $X$ are a part of the ${\bf 27}$-plet in the $E_6$.
To be more general, however,  here we allow $n$ to take a different value.
%To break the extra U(1)$_X$, we introduce a SM singlet Higgs $X$ with U(1)$_\psi$ charge $n$ and assume that it obtains a VEV.
The mass of $\psi$ is generated through 
\begin{equation}
	-\mathcal L = 
	\frac{1}{2} M  \lrfp{\la X^* \ra}{M}{8/n}\psi\psi + {\rm h.c.}
\end{equation}
Thus the possible value of $n$ is $n = 1,2,4,8$.
Among these, $n=8$ requires a very small VEV of $X$ for generating small $\psi$ mass.
In the following we consider the cases of $n=4$ and $n=1$.
The situation for $n=2$ lies between these two cases.
After $X$ develops a non-zero VEV, $\psi$ obtains a mass of
\begin{equation}
	m_\psi \;\simeq \;\begin{cases}
	\displaystyle 7\,{\rm keV}\left( \frac{\langle X\rangle}{4\times 10^6\,{\rm GeV}} \right)^2\left( \frac{M_P}{M} \right) &{\rm for~}n=4\\
	\displaystyle 7\,{\rm keV}\left( \frac{\langle X\rangle}{2.8\times 10^{15}\,{\rm GeV}} \right)^8\left( \frac{M_P}{M} \right)^7 &{\rm for~}n=1.
	\end{cases}
\end{equation}
This light fermion may act as the DM in the universe.\footnote{
	We can also write down the $\psi$-$\nu_{\rm R}$ mixing term, $\mathcal L\sim  (X^*)^{4/n}\phi^*\psi \nu_{\rm R}/M^{4/n} + {\rm h.c.}$,
	but inclusion of this term does not much affect the result.
}
The other extra fermions, $\psi_{\bf 5}$ and $\psi_{\bf \bar 5}$, obtain the Dirac mass through
\begin{equation}
	-\mathcal L \simeq
	 M \lrfp{\la X \ra}{M}{4/n} \psi_{\bf 5}\psi_{\bf \bar 5} + {\rm h.c.}.
\end{equation}
Thus its mass is given by $\sqrt{m_\psi M} \sim 4\times 10^6$\, GeV for $m_\psi = 7$\,keV and 
$M = M_P$. Note that the mass is independent of $n$ for the given $m_\psi$.

The mixing between active neutrinos and $\psi$ allows $\psi$ to decay into the neutrino and photon,
similarly to the sterile neutrinos.
The mixing arises from the operator
\begin{equation}
	-\mathcal L = c \lrf{\la \phi \ra}{M} \lrfp{\la X^* \ra}{M}{4/n} H L \psi + {\rm h.c.},
\end{equation}
where $H$ is the SM Higgs doublet and $L$ is the lepton doublet and $c$ is a numerical constant of order unity.
The mixing angle between $\psi$ and active neutrinos, $\theta$, can be evaluated as
\begin{align}
	\theta & \simeq c \lrf{\la \phi \ra}{M} \lrf{v}{M}  \lrfp{\la X \ra}{M}{-4/n}
	\simeq \frac{c\langle \phi\rangle  v}{m_\psi^{1/2} M^{3/2}}\\
	&\simeq 9\times 10^{-7} c
	\left( \frac{\langle \phi\rangle}{5\times 10^{16}\,{\rm GeV}} \right)
	\left( \frac{7\,{\rm keV}}{m_\psi} \right)^{1/2}
	\left( \frac{M_P}{M} \right)^{3/2},
\end{align}
where $v=174\,$GeV is the VEV of the SM Higgs and we have omitted $\tan \beta$ for simplicity.
Most important, this value of the mixing is close to the value required to explain the excess of the $3.5$\,keV X-ray line:
$\theta \sim 4 \times 10^{-6}$~\cite{Boyarsky:2014jta}. Note that the mixing is independent of $n$
for the given mass of $\psi$. Thus, the required longevity (\ref{tau}) is automatic in our model.

The abundance of $\psi$ depends on thermal history of the universe.
For high enough reheating temperature after inflation, $\psi$ is thermalized though U(1)$_X$ gauge interactions.
The freezeout temperature of $\psi$ is estimated to be $T_f\sim (\langle X\rangle^4/M_P)^{1/3} \sim \mathcal O(100)\,$GeV for $\langle X\rangle \sim 10^6$\,GeV $(n=4)$ 
and $T_f \sim 10^{14}\,$GeV for $\langle X\rangle \sim 10^{15}$\,GeV $(n=1)$
for the U(1)$_X$ gauge coupling constant of $\mathcal O(1)$.
Assuming that three generations $\psi$ all have equal masses for simplicity, we obtain the relic abundance as
\begin{equation}
	\frac{\rho_\psi}{s} \simeq 8\times 10^{-8}\,{\rm GeV} \left( \frac{m_\psi}{7\,{\rm keV}} \right)
	\left( \frac{106.75}{g_{*s}(T_f)} \right),
\end{equation}
if the reheating temperature is higher than the freeze-out temperature, i.e., if they are thermalized.
Here $g_{*s}(T_f)$ counts the relativistic degrees of freedom at the freeze-out. 
If $\psi$ particles are thermalized, we need a dilution factor of $\sim 200$ in order to obtain correct relic DM abundance.
Suppose that the reheating temperature is so high that the extra heavy fermions, $\psi_{\bf 5}$ and $\psi_{\bf {\bar 5}}$, 
are also efficiently produced. This is indeed the case of $n=1$ as the freeze-out temperature is higher than the mass of 
$\psi_{\bf 5}$ and $\psi_{\bf {\bar 5}}$. They are stable in the presence of the exact $Z_2$ symmetry, but they decay into SM particles
if the $Z_2$ parity is weakly violated. For a certain breaking of the $Z_2$ symmetry, the decay could produce the required
entropy production to obtain the right DM abundance of $\psi$~\cite{Fujii:2002fv}.
Alternatively, if the reheating temperature is slightly lower than the freeze-out temperature, 
a correct amount of DM can be produced.

%%%%%%%%%%%%%%%%%%%%%%%%%%%%%%%%%%%%%%%%%%%%
\section*{Acknowledgments}
%%%%%%%%%%%%%%%%%%%%%%%%%%%%%%%%%%%%%%%%%%%%

This work was supported by the Grant-in-Aid for Scientific Research on
Innovative Areas (No.23104008 [FT]),  JSPS Grant-in-Aid for
Young Scientists (B) (No. 26800121 [KN], No.24740135) [FT]), and Inoue Foundation for Science [FT].  
This work was also supported by World Premier International Center Initiative (WPI Program), MEXT, Japan.

%%%%%%%%%%%%%%%%%%%%%%%%%%%%%%%%%%%%%%%%%%%%

\end{document}